\documentclass[RNAAS]{aastex63}
\usepackage{graphicx,epsfig,epsf}
\usepackage{threeparttable}
\usepackage{multirow}
\usepackage{graphicx}
\usepackage{times}
\usepackage{natbib}
\usepackage{amssymb,amsmath}
\usepackage{indentfirst}
\usepackage{array}
\usepackage{booktabs} 
\usepackage{threeparttable}
\usepackage{rotating}
\usepackage{float}
\usepackage{color}
\usepackage{times}
\usepackage{url}
\usepackage{subfigure}

\graphicspath{{./}{figures/}}
\begin{document}

\title{Re-estimating the Spin Parameter of the Black Hole in Cygnus X-1}

\correspondingauthor{Xueshan~Zhao,Lijun~Gou}
\email{xszhao@nao.cas.cn,lgou@nao.cas.cn}

\author{Xueshan~Zhao}
\affiliation{Key Laboratory for Computational Astrophysics, National Astronomical Observatories, Chinese Academy of Sciences,
Datun Road A20, Beijing 100012, People’s Republic of China; xszhao@nao.cas.cn, lgou@nao.cas.cn}
\affiliation{School of Astronomy and Space Sciences, University of Chinese Academy of Sciences, Datun Road A20, Beijing 100049, People’s Republic of China}

\author{Lijun~Gou}
\affiliation{Key Laboratory for Computational Astrophysics, National Astronomical Observatories, Chinese Academy of Sciences,
Datun Road A20, Beijing 100012, People’s Republic of China; xszhao@nao.cas.cn, lgou@nao.cas.cn}
\affiliation{School of Astronomy and Space Sciences, University of Chinese Academy of Sciences, Datun Road A20, Beijing 100049, People’s Republic of China}

\author{Yanting~Dong}
\affiliation{Key Laboratory for Computational Astrophysics, National Astronomical Observatories, Chinese Academy of Sciences,
Datun Road A20, Beijing 100012, People’s Republic of China; xszhao@nao.cas.cn, lgou@nao.cas.cn}
\affiliation{School of Astronomy and Space Sciences, University of Chinese Academy of Sciences, Datun Road A20, Beijing 100049, People’s Republic of China}

\author{Xueying~Zheng}
\affiliation{Key Laboratory for Computational Astrophysics, National Astronomical Observatories, Chinese Academy of Sciences,
Datun Road A20, Beijing 100012, People’s Republic of China; xszhao@nao.cas.cn, lgou@nao.cas.cn}
\affiliation{School of Astronomy and Space Sciences, University of Chinese Academy of Sciences, Datun Road A20, Beijing 100049, People’s Republic of China}

\author{James F. Steiner}
\affiliation{Harvard-Smithsonian Center for Astrophysics, Cambridge, MA 02138, USA}

\author{James C.A. Miller-Jones}
\affiliation{International Centre for Radio Astronomy Research – Curtin University, GPO Box U1987, Perth, WA 6845, Australia}

\author{Arash Bahramian}
\affiliation{International Centre for Radio Astronomy Research – Curtin University, GPO Box U1987, Perth, WA 6845, Australia}

\author{Jerome A. Orosz}
\affiliation{Astronomy Department, San Diego State University, 5500 Campanile Drive, San Diego, CA 92182-1221, USA}

\author{Ye~Feng}
\affiliation{Key Laboratory for Computational Astrophysics, National Astronomical Observatories, Chinese Academy of Sciences,
Datun Road A20, Beijing 100012, People’s Republic of China; xszhao@nao.cas.cn, lgou@nao.cas.cn}
\affiliation{School of Astronomy and Space Sciences, University of Chinese Academy of Sciences, Datun Road A20, Beijing 100049, People’s Republic of China}


\begin{abstract}
  Cygnus X-1 is a well-studied persistent black hole X-ray binary. Recently, the three parameters needed to estimate the black hole spin of this system, namely the black hole mass $M$, the orbital inclination $i$ and the source distance $D$, have been updated. In this work we redetermine the spin parameter using the continuum-fitting technique for those updated parameter values.  Based on the assumption that the spin axis of the black hole is aligned with the orbital plane, we fit the thermal disk component to a fully relativistic thin accretion disk model. The error in the spin estimate arising from the combined observational uncertainties is obtained via Monte Carlo (MC) simulations. We demonstrate that, without considering the counteracting torque effect, the new spin parameter is constrained to be a$_* > 0.9985$ (3$\sigma$), which confirms that the spin of the black hole in Cygnus X-1 is extreme.
\end{abstract}


\section{introduction} 
\label{intro}

The spin parameter is a fundamental physical property of a black hole, which can be estimated observationally using one of two widely-used methods;
the iron-line fitting method (\citealt{Fabian+etal+1989}, \citealt{Reynolds+etal+2003}) and the continuum-fitting method (\citealt{Zhang+etal+1997}). In this work we use the continuum-fitting method to estimate the dimensionless spin parameter $a_{*}$ ($a_{*} \equiv a/M \equiv cJ/GM^2$, where $M$ and $J$ indicate the BH mass and angular momentum, respectively) of the stellar-mass black hole Cygnus X-1. Up to now, this method has been extensively employed in roughly a dozen stellar-mass black holes: 4U 1543-47 ($a_{*}=0.43^{+0.22} _{-0.31}$, 90\% confidence,~\citealt{Morningstar+etal+2014}), A0620-00 ($a_{*} = 0.12 \pm 0.19$, 1$\sigma$, ~\citealt{Gou+etal+2010}), Cygnus X-1 ($a_{*} > 0.983$, 3$\sigma$,~\citealt{Gou+etal+2014}), GRO J1655-40 ($a_{*} = 0.70 \pm 0.10$, 1$\sigma$,~\citealt{Shafee+etal+2006}), GRS 1915+105 ($a_{*} > 0.95$, 3$\sigma$,~\citealt{McClintock+etal+2006}), LMC X-1 ($a_{*}=0.92^{+0.05} _{-0.07}$, 1$\sigma$,~\citealt{Gou+etal+2009}), LMC X-3 ($a_{*}=0.25^{+0.13} _{-0.16}$, 1$\sigma$,~\citealt{Steiner+etal+2014}), M33 X-7 ($a_{*} = 0.84 \pm 0.05$, 1$\sigma$,~\citealt{Liu+etal+2010}), Nova Muscae 1991 ($a_{*}=0.63^{+0.16} _{-0.19}$, 1$\sigma$,~\citealt{Chen+etal+2016}), XTE J1550-564 ($a_{*} = 0.34^{+0.20} _{-0.28}$, 1$\sigma$,~\citealt{Steiner+etal+2011}), IC10 X-1 ($a_{*} \gtrsim 0.8$, 90\% confidence,~\citealt{Steiner+etal+2016}), GX 339-4 ($a_* < 0.9$, \citealt{Kolehmainen+etal+2010}), H1743-322 ($a_{*} = 0.2 \pm 0.3$, 1$\sigma$,~\citealt{Steiner+etal+2012}).

In applying this method, the key fitting parameter is the inner radius of the thin accretion disk $R_{\rm in}$, which can be determined by fitting the X-ray continuum spectrum to the Novikov-Thorne thin disk model (\citealt{Novikov+etal+1973}). $R_{\rm in}$ is considered as equivalent to the radius of the innermost stable circular orbit $R_{\rm ISCO}$ (\citealt{Shafee+etal+2008}, \citealt{Penna+etal+2010}, \citealt{Noble+etal+2010}, \citealt{Kulkarni+etal+2011}). In addition, $R_{\rm ISCO}$ is a monotonic function of the dimensionless spin parameter $a_{*}$, decreasing from 6 $GM/c^{2}$ to 1 $GM/c^{2}$ as the spin increases from $a_{*}$ = 0 to $a_{*}$ = 1. Therefore, given the inner radius, one can constrain the value of $a_{*}$ through this simple monotonic relationship~\citep{Bardeen+etal+1972}.

In order to make sure the inner radius of the disk has reached the ISCO, the high/soft state spectra dominated by the accretion disk component are usually rigorously selected to avoid confusion introduced by the strongly Comptonized component (\citealt{Shafee+etal+2006}, \citealt{McClintock+etal+2006}, \citealt{Liu+etal+2008}). The method requires a thin disk with a moderate luminosity, which means that the ratio of the disk thickness and the local disk radius should satisfy $H/R < 0.03$, and the bolometric Eddington-scaled luminosity $l = L(a_*,\dot M)/L_{\rm Edd} < 0.3$ (\citealt{McClintock+etal+2006}, \citealt{Zhu+etal+2012}, \citealt{McClintock+etal+2014}). However, in practice, the high/soft spectra are found to be difficult to select for some sources, such as Cygnus X-1, which has never reached a canonical high/soft state according to its standard definition (\citealt{Gou+etal+2011}). Therefore, ~\cite{ste+2009b} proposed an alternative approach to select the spectra, namely that the scattering fraction $f_{\rm sc}\lesssim25\%$. The scattering fraction refers to the fraction of the thermal seed photons that are scattered into the power-law component, and is quantified via the parameter $f_{\rm sc}$ in the Comptonization model SIMPL (\citealt{ste+2009a}). The maximum value of the fraction is assumed to be unity. This approach has been successfully applied to several black hole systems (\citealt{Gou+etal+2009}, \citealt{Steiner+etal+2011}, \citealt{Chen+etal+2016}), and it was found that some steep power-law (SPL) state spectra could also be used for the spin measurement, in addition to the standard high/soft state spectra. This criterion thereby allowed the spectral sample to be extended beyond the pure thermal state. The advantage of this approach is not only that all the selected spectra give the roughly same inner radius (i.e., the spin parameters remain quite stable both in the thermal state and the more strongly Comptonized SPL state), but also that the approach can be readily implemented (\citealt{Gou+etal+2009}, \citealt{Steiner+etal+2010}, \citealt{Chen+etal+2016}).

The success of the continuum-fitting method relies on having accurate values of three input dynamical parameters beforehand, namely the black hole mass $M$, the disk inclination $i$ and the source distance $D$. The spin error introduced by the combined observational uncertainties of $M$, $i$ and $D$, which together dominate the error budget in spin, is typically estimated via Monte Carlo simulations~(\citealt{Liu+etal+2008}, \citealt{Gou+etal+2011},~\citealt{McClintock+etal+2014}).

Cygnus X-1 is the first confirmed black hole candidate system, and is one of the most famous persistent black hole X-ray binaries, whose dynamical parameters have been well-estimated. A systematic study of the system was carried out by~\cite{Reid+etal+2011},~\cite{Orosz+etal+2011}, and~\cite{Gou+etal+2014}.~\cite{Reid+etal+2011} estimated the distance using trigonometric parallax, providing a precise, model-independent distance measurement for the system. Afterwards, using the distance as one of the constraints,~\cite{Orosz+etal+2011} fit the optical photometric light curves and radial velocity curves to determine the mass of the black hole and the inclination angle of the system. With the system parameters in hand, \cite{Gou+etal+2011} firstly found a near-extreme value of the spin, $a_{*} > 0.95$ (3$\sigma$), via the continuum-fitting method, and \cite{Gou+etal+2014} later confirmed this extreme result but with a more stringent lower limit $a_{*} > 0.983$ (3$\sigma$). 

Recently, \cite{Miller-Jones+etal+2020} made a new set of six daily orbital-phase resolved {\it VLBA} observations, which revealed an orbital phase dependence in the measured astrometric positions. Hence, free-free absorption in the stellar wind will affect the astrometry by changing the position of the optical depth unity surface seen as the radio core, which if unaccounted for can affect the measured parallax signal. They therefore combined the archival data of \cite{Reid+etal+2011} with new {\it VLBA} data, and conducted their astrometric fit only in the dimension perpendicular to the jet axis on the plane of the sky, thereby negating the impact of the stellar wind absorption. This gave a refined measurement of the distance, of $D$ = 2.22 $^{+0.18}_{-0.17}$ kpc. The updated distance measurement required a revision of the optical modelling, leading to new values for both the black hole mass and orbital inclination, of $M$ = 21.2 $\pm$ 2.2 $M_{\odot}$, $i$ = 27.51 $^{\circ}$ $^{+0.77}_{-0.57}$ , with all uncertainties quoted at 1$\sigma$. Using these up-to-date values, we are able to refit the X-ray spectra used in~\cite{Gou+etal+2011} and~\cite{Gou+etal+2014}, and refine the constraints on the black hole spin. The updated results are presented in this paper. 

This paper is organized as follows. In Section~2, we discuss our observations. In Section~3, we describe our detailed spectral analysis. A discussion is presented in Section~4. Finally in Section~5, we summarize our results and give a conclusion.

\section{observations}
\label{obse}
Because Cygnus X-1 has never reached the classically-defined high/soft state, we used the scattering fraction criterion outlined in Section~\ref{intro} to select the spectra. We found several spectra that could be used for the spin measurement, which were originally presented by ~\cite{Gou+etal+2011, Gou+etal+2014} (hereafter GOU11 and GOU14). In this paper, we used exactly the same spectra for our analysis. The log of these observations is presented in Table~1 for convenience. 

The 1996 archival {\it ASCA}\&{\it RXTE} spectrum (hereafter SP1) was discussed in GOU11. We disregarded the other two spectra analyzed in GOU11 because they have $f_{\rm sc}$ above $25\%$ ($30.5\% \pm 1.2\%$ and $30.6\% \pm 0.6\%$, respectively), which is the upper limit for reliable measurements using the continuum-fitting method. Details of the 2011 archival {\it Chandra}\&{\it RXTE} spectra (hereafter, SP2-SP6) were presented in GOU14. These crucial soft state spectra were all extracted from a single observation, which was made on Jan. 6th, 2011. 

Following GOU11 and GOU14, we also corrected for the effective area of the PCA using the Crab Nebula as a calibration source (\citealt{Toor+etal+1974}). We applied several correction factors for {\it RXTE} (see \citealt{Steiner+etal+2010} for specific definitions of these correction factors), and the respective values for SP1 and SP2-SP6 are (as described in GOU11 and GOU14): the normalization correction factors $C_{\rm TS}$ (the ratio of the observed normalization to that of \citealt{Toor+etal+1974}) of 1.123/1.128; the power-law slope correction factors $\triangle\Gamma _{\rm TS}$ (the difference between the observed value of $\Gamma$ and that of \citealt{Toor+etal+1974}) of 0.023/0.022; and the dead time correction factors (the normalization correction factors for detector dead time) of 1.048/1.029.

For {\it Chandra} observation in SP2-SP6, we rebinned the data to achieve a minimum of 200 counts in each new bin. For {\it ASCA} and {\it RXTE} observations, we rebinned the data with at least 25 counts within each bin.

Finally, we included systematic errors to account for the uncertainties in the instrumental responses: 0.5\% for {\it RXTE} and and 1\% for {\it ASCA}.\footnote{To achieve an acceptable fit to the PCA spectra, the standard prescription recommended the addition of a 0.5\%--1\% systematic uncertainty before fitting the spectra, as we did here. However, \cite{Gar+etal+2014} proposed a new empirical calibration model named PCACORR, based on the fitting of hundreds of {\it RXTE} PCA spectra of the Crab. They found that the quality of the spectra could be significantly improved and required only a 0.1\% systematic uncertainty. This was tested in \cite{Gar+etal+2014} and \cite{Gar+etal+2016}. However, for consistency with the previous measurements in GOU11 and GOU14, we did not apply the PCACORR tool to the spectra. However, we did one experiment, applying the PCACORR tool to all the PCA spectra and refitting the spectra, and we found the results to be almost the same, with negligible changes.}


\section{spectral analysis}
\label{spec}
We performed our data analysis and spectral fitting using the software package XSPEC version 12.10.1\footnote{XSPEC is available at http://heasarc.gsfc.nasa.gov/xanadu/xspec/}. The fit statistic in use is chi-square $\chi^2$. And the photoelectric absorption cross-sections in use is based on \cite{Verner+etal+1996}. During the initial fitting process, the three key input parameters $M$, $i$ and $D$ are fixed at their newly-obtained values given in Section~\ref{intro}.
\subsection{Our Adopted Model}
Three typical components might be included in the spectrum of a black hole X-ray binary, namely the thermal, power-law and reflection components (refer to Figure 2 in GOU11 for a schematic illustration of these different components). 

The model we employed in this paper is the most advanced and physically realistic model used in GOU11 and GOU14. The structure of the model, naming all the components that it comprises, with a concise description of each one, are as follows (the detailed parameter configuration is the same as GOU11 and GOU14, unless otherwise indicated):

\vspace{-3mm}

\begin{eqnarray*}
{\rm CRABCOR*CONSTANT*TBABS~[SIMPLR \otimes KERRBB{\small 2}  +KERRDISK+~KERRCONV\otimes (IREFLECT \otimes SIMPLC)]}
\end{eqnarray*}

{\it Thermal component}: The thermal radiation is emitted from the accretion disk. {\sc kerrbb{\small 2}}~\citep{McClintock+etal+2006} is a fully relativistic thin disk model, including the effects of the returning radiation and the limb darkening. This model reads in and uses a pair of look-up tables for the spectral hardening factor $f$, computed using {\sc bhspec}~(\citealt{Davis+etal+2005},~\citealt{Davis+etal+2006}) and {\sc kerrbb} (\citealt{Li+etal+2005}), corresponding to two representative values of the viscosity parameter: $\alpha$ = 0.01 and $\alpha$ = 0.1 (here we adopt $\alpha$ = 0.1). {\sc kerrbb{\small 2}} only has two principal fit parameters: the dimensionless spin parameter a$_{*}$, and the mass accretion rate $\dot{M}$.

{\it Power-law Component}: The thermal seed photons supplied by the accretion disk are inverse-Compton scattered into the power-law component by hot electrons in the corona. The convolution model {\sc simplr}, a modified version of {\sc simpl} (\citealt{ste+2009a}), is an advanced empirical Comptonization model including three main fit parameters: the photon power-law index $\Gamma$, the scattered fraction $f_{\rm sc}$ and the reflection constant parameter $\chi$. As we mentioned, we adopt f$_{\rm sc} \lesssim 25\%$ as our spectrum selection standard. 

{\it Reflection Component}: A fraction of the scattered seed photons are emitted downward toward the disk, illuminating the disk and generating the reflection component. {\sc kerrdisk} + {\sc kerrconv}$\otimes$ ({\sc ireflect} $\otimes$ {\sc simplc)]} models the entire reflection component including the Compton hump, the absorption edge and the emission lines. {\sc simplc} represents the part of the power-law component that strikes the disk (\citealt{ste+2009a}), which is regarded as the input spectrum for the convolution model {\sc ireflect} (\citealt{Magdziarz+etal+1995}). Photons emitted near the black hole suffer various relativistic effects, leading to a blurring of the observed spectrum. {\scshape kerrconv} is utilized with the aim of smearing the spectrum~\citep{Brenneman+etal+2006}. {\scshape kerrdisk} is a model for the iron emission line with the spin allowed to be a free parameter~\citep{Brenneman+etal+2006}. Here it should be noted that the effect on the spin measurements of using different reflection models is almost negligible, as demonstrated in section 5.3 of GOU11.

In addition, we use three multiplicative models: (1) {\sc crabcor} is used for correcting calibration deviations of {\it RXTE}. (2) {\sc constant} is an energy-independent multiplicative factor in order to reconcile calibration discrepancies between different detectors. (3) {\sc tbabs} is an ISM absorption model (\citealt{Wilms+etal+2000}).

\subsection{Fit Results} 
Our model provides a good fit to all six spectra, with reduced $\chi^2$ values varying from 1.12 to 1.49. The fit parameters are shown in Table~2. A representative plot of the unfolded spectrum is given in Figure~1. It is noticed that there is a larger residual between 8\,keV and 15\,keV, which will be discussed in detail later, and should not significantly affect the spin results.

As a caveat, in all six cases, the values of $a_{*}$ reached the maximum value of the XSPEC model {\sc kerrbb{\small 2}}, 0.9999. This value is larger than the physical upper limit of a$_{*}$, which we will discuss in the next section. The preliminary error in $a_{*}$ computed with XSPEC and listed in Table~2 is due to counting statistics, while the final adopted error due to the combined uncertainties in $D$, $M$, and $i$ is acquired via our Monte Carlo analysis.

Requiring the spectrum to be dominated by accretion disk component for reliable application of the continuum-fitting method, we adopt $f_{\rm sc} \lesssim 25\%$ as the selection standard. The values of this parameter in our analysis, 24.7\% $\pm$ 0.5\%, 10.6\% $\pm$ 1.6\%, 15.6\% $\pm$ 1.2\%, 21.1\% $\pm$ 1.0\%, 16.3\% $\pm$ 0.9\%, 9.3\% $\pm$ 0.2\%, satisfy this standard.

It is equally essential to select spectra with bolometric Eddington-scaled luminosities $l < 0.3$ (\citealt{McClintock+etal+2006}, \citealt{Zhu+etal+2012}, \citealt{McClintock+etal+2014}), in order to ensure a geometrically thin disk. This parameter is quite stable and low in our analysis, ranging from 0.01 to 0.03, which meets this criteria without difficulty.

\subsection{Error Analysis}

Previous work on estimating $a_{*}$ had found that the combined observational uncertainties in $M$, $i$ and $D$ dominate the error budget~(GOU11,~\citealt{McClintock+etal+2014}). This part of the error is usually calculated through Monte Carlo simulations (\citealt{Liu+etal+2008}, \citealt{Gou+etal+2011},~\citealt{McClintock+etal+2014}). In order to do so, for each individual spectrum, the procedure is as follows: (1) Assuming that each one of these three parameters is independent and Gaussian distributed, we generate 3000 sets of ($M$, $i$, $D$). (2) Then we calculate look-up tables of $f$ for these parameter sets. (3) Finally we fit the spectrum to our adopted model to determine $a_{*}$. With 3000 simulations run, we obtain the histogram of $a_{*}$.

After repeating the procedure expressed above for all six spectra, we obtain six histograms of $a_{*}$. The histogram for each spectrum is shown in Figure~\ref{fig2}, and the summed histogram is plotted in Figure~\ref{fig3}. The spin value in SP2 is lowest, therefore we report the final result of $a_{*} > 0.9985$ $(3\sigma)$ based on a conservative estimation from this spectrum.

\section{discussion}

\subsection{Comparison with the Thorne Limit}

In this paper, we have obtained a much more stringent limit on the spin of the black hole in Cygnus X-1 than found from previous work: $a_{*} > 0.9985$ at 3$\sigma$ level of confidence. This very high spin parameter confirms that Cygnus X-1 hosts an extreme Kerr black hole.

However, \cite{Thorne+1974} pointed out that the maximum expected spin once the absorption of hot photons from the disk is taken into account is 0.998 (hereafter, the Thorne limit), as the absorption will produce an counteracting torque when the spin is extremely high. We note that our spin constraint is slightly higher than the Thorne limit. This is mainly because the upper limit in our model {\sc kerrbb{\small 2}} is set at 0.9999, and our obtained spin parameter is based purely on the theoretical model.  

It should be noted that although the Thorne limit has generally been adopted by the community as a reasonable guess for an astrophysical upper limit, there is still some debate as to the exact value of the upper limit. \cite{sad+etal+2011} suggested that black holes with super-critical accretion flows could reach an equilibrium spin value up to 0.9994, which exceeds the Thorne limit. They showed that the equilibrium spin value depends strongly on the assumed value of the viscosity parameter, and also proved that for high accretion rates the impact of captured radiation on spin evolution is negligible.  Cygnus X-1 is not believed to be in the super-critical regime, and hence these considerations may not apply. However, a number of other factors might affect the precise spin measurement, and they are discussed in the following sections.  Regardless, the spin of Cygnus X-1 appears to be very high, as found previously by both continuum fitting (GOU11, GOU14) and iron line fitting studies (\citealt{Fabian+etal+2012}, \citealt{Tomsick+etal+2014}, \citealt{Parker+etal+2015}, \citealt{Duro+etal+2016}, \citealt{Walton+etal+2016}, \citealt{Tomsick+etal+2018}).

\subsection{Effect of a Possible Spin-orbit Misalignment}

Figure~\ref{fig4} shows the effect on $a_{*}$ of separately varying the input parameters $M$, $i$, and $D$. $a_{*}$ takes the extreme value of 0.9999 when all three parameters are fixed on their newly-obtained, best-fitting values. 

In applying the continuum-fitting method, we assume that the binary orbital angular momentum vector is aligned with the spin axis of the black hole. Cygnus X-1 is suggested to have formed with little to no natal kick (\citealt{Mirabel+etal+2003}, \citealt{Rao+etal+2019}), so the spin axis of the black hole should still be aligned with the orbital angular momentum vector (as discussed by \cite{Miller-Jones+etal+2020}, who estimated a likely maximum misalignment angle of order $10^\circ$). 

However, there is still controversy about this assumption. Some previous work on fitting the reflection component of Cygnus X-1 suggested a different inner disk inclination angle compared to the binary orbital inclination, which implies a possible spin-orbit misalignment (also suggested in some other black hole binary systems, \citealt{Fragos+etal+2010}, \citealt{Dong+etal+2020}). \cite{Tomsick+etal+2014} evaluated the lower limit of this misalignment to be $13^\circ$. \cite{Walton+etal+2016} estimated the magnitude of a disk warp to be about $10^\circ-15^\circ$. However, these fits often require very high iron abundances, and \cite{Tomsick+etal+2018} recently found that a much lower disk inclination of $<$ 20$^\circ$ could be accommodated if the electron density was free to vary, even suggesting that a misalignment might not even be required in this scenario.  Thus, while the possibility of a misalignment should be considered, this scenario would appear to remain under debate.

Nonetheless, we investigated this possibility, and as shown in Figure~\ref{fig4}(a), the value of $a_{*}$ monotonically decreases with increasing $i$. We find that even given a misalignment angle of $15^\circ$ (equivalent to $i$ = $42.47^\circ$), the spin value is 0.9696 (see Case 1 in Table~2), which is still very high.

\subsection{Effect of the Metallicity}

The spectral hardening factor $f$ in {\sc kerrbb{\small 2}} is pre-calculated with a table model {\sc bhspec}, which is related to the metallicity $Z$. Our fit results are all given for $Z$ = 1 $Z_\odot$. However, the metallicity of Cygnus X-1 is slightly super-solar (\citealt{Shimanskii+etal+2012}), as also implied by the fitted super-solar values of Fe abundance in our reflection model {\sc ireflect}.  

 Unfortunately, the current model {\sc bhspec} only contains two possible metallicity cases ($Z$ = 1 $Z_\odot$ and $Z$ = 0.1 Z$_\odot$), and does not yet have an option for super-solar abundance. However, if the metallicity increases, the value of the spin increases slightly. In addition, substantial changes in metallicity only produce a negligible effect on spin.  For example, as stated in~\citealt{Gou+etal+2009}, the spin of the black hole in LMC X-1 increases from 0.937 $\pm$ 0.020 to 0.938 $\pm$ 0.020 if Z increases from 0.1 Z$_\odot$ to 1 Z$_\odot$.

\subsection{Effect of the Viscosity Parameter}

Throughout this work, in order to get a more conservative constraint, we fixed the viscosity parameter $\alpha$ to 0.1, because $a_{*}$ increases slightly as $\alpha$ decreases (\citealt{Gou+etal+2009}). However, we reanalyzed our data using $\alpha$ = 0.01, and found that the results remain unchanged, as they have already reached their maximum value of 0.9999.

\subsection{Effect of a Finite Disk Height}

The model utilized in the continuum-fitting method is the general relativistic Novikov-Thorne model that assumes a razor-thin disk. However, a real accretion disk will have finite thickness, which will affect measurements of the black hole spin. \cite{Kulkarni+etal+2011} evaluated the error introduced by the deviations from the Novikov-Thorne model. They computed disk spectra obtained from three-dimensional general relativistic magnetohydrodynamic (GRMHD) simulations, and then fitted these simulated spectra with {\scshape kerrbb}. By comparing the spin results with the spin used in the GRMHD simulations, they concluded that the model errors become obvious only at low spins and high inclinations. In the case of $i$ = 30$^\circ$ and a$_*$ = 0.98 (close to the values estimated for Cygnus X-1), the spin calculated by fitting the simulated spectra was 0.985 $\pm$ 0.001 and the value of $\triangle$ $a_*$ is 0.005. Moreover, they might have overestimated the error due to the deviation from the Novikov-Thorne model because their simulation results are for a thicker disk with a bolometric Eddington-scaled luminosity $l$ = 0.4--0.7, while the bolometric Eddington-scaled luminosity of Cygnus X-1 in this paper is only 0.02 (as stated in Section 6 of GOU11).

%




\subsection{The Residual Structure between 8–15 keV}

We notice that there is a larger residual between 8\,keV and 15\,keV, which is not fitted with the current model.  However, it should be emphasized that the spin parameter is determined from the disk thermal component (which dominates the spectra, peaks around 1.5 keV, and is concentrated below 6 keV; see Figure 3 of GOU11 for reference), and thus the residual between 8 and 15\,keV should not significantly affect the spin results.

Previous work has proposed many theories to explain similar residual structures. \cite{Fabian+etal+2020} found an 7–9 keV excess emission component during the soft state in spectra of MAXI J1820+070. They explained this additional component by including another blackbody whose temperature is slightly higher than that of the disk itself. Given that the higher temperature, the additional emission is linked to the innermost region of the disk and with the start of the plunging region. In this region, the density is still high enough to heat the gas, possibly emitting blackbody radiation. We utilized an addition model {\scshape BBODY} to describe this component, we find that the fit is poor ($>2$), and the spin could not be determined certainly.

Another possibility is that the observed residual structure is an additional Comptonization component. In our fits, we only utilized Comptonization models which have a pure non-thermal power-law electron distribution. However, some previous work of the fast time variability seen in the high/soft state prefer a stratified Comptonization spectrum (\citealt{Grinberg+etal+2014}). It is likely that there exist hybrid thermal/non-thermal electron distributions, resulting in an additional low temperature thermal comptonization component (\citealt{Gierli+etal+1999}, \citealt{Zdziarski+etal+2002}, \citealt{Kawano+etal+2017}). This component then produce a higher energy power-law tail, which would also lead us to overestimate the black hole spin. For instance, as mentioned in \cite{Kawano+etal+2017}, the spin of the black hole in Cygnus X-1 is 0.95 $\pm$ 0.01 for a simple non-thermal Comptonization; and decreases to 0.80 $^{+0.08}_{-0.30}$ for hybrid thermal/non-thermal Comptonization. A similar residual structure between 5–10 keV was found in the spectra of the black hole candidate MAXI J1828-249 in a possible intermediate state (\citealt{Oda+etal+2019}). They modeled this excess with a Comptonization of disk photons by thermal electrons with a relatively low temperature and inferred it could be generated in a region where the disk inner edge intrudes into the hot inner flow, or a Comptonizing region above the disk surface. We used an addition model {\scshape comptt} to describe the Comptonization of soft photons, and reflected both two Comptonization components. The model is reconstructed as follows:

\vspace{-3mm}
\begin{eqnarray*}
{\rm CRABCOR*CONSTANT*TBABS~[KERRBB{\small 2}+SIMPL \otimes COMPTT}\\
{\rm +KERRDISK+~KERRCONV\otimes (IREFLECT \otimes SIMPL \otimes COMPTT)]}
\end{eqnarray*}

The fit parameters are listed in Table~3. For SP1, the fit is better than former model with reduced $\chi^2$ reducing from 1.49 to 0.94. The spin parameter $a_*$ decreases from 0.9999 to $0.9947 \pm 0.0014$. The photon power-law index decreases from $2.39 \pm 0.01$ to $1.99 \pm 0.01$. The scattered fraction $f_{\rm sc}$ is larger ranging from $0.247 \pm 0.005$ to $0.399 \pm 0.007$. For SP2-SP6, the values of $a_*$  are unaffected, which are still at 0.9999. However, the errors quoted here (estimated in XSPEC) are obviously larger. Besides, for these five spectra, this model can't constrain the parameters of {\sc comptt} well probably because the component is weak. The scattered fraction $f_{\rm sc}$ are somehow increasing from $\sim 0.15$ to $\sim 0.40$. For SP5, the fit is slightly worse with reduced $\chi^2$ increasing from 1.25 to 1.27. While for SP2, SP3, SP4 and SP6, the fit is better with reduced $\chi^2$ ranging from 1.12, 1.15, 1.32, 1.24 to 0.93, 1.02, 0.94, 1.12, respectively. We also try the same setup of \cite{Kawano+etal+2017}, fixing $T_{\rm 0}$, $kT$ and $\tau$ at 0.44 keV, 3.4 keV and 1.66, respectively. Details are shown in Case 1 of Table~3. The spin parameter $a_*$ is still pegged at 0.9999. 

Whether using a hybrid thermal/non-thermal Comptonization model or a single non-thermal Comptonization model, we all get an extremely high spin. So we can conclude that including {\sc comptt} is not having the same effect as in \cite{Kawano+etal+2017}. However, it is interesting to note that the reduced $\chi^2$ decreases significantly (e.g., for SP1 it goes to 0.94 from 1.49), which means that adding this Comptonization model is having a large impact on the high energy Compton component rather than on the disk which is the foundation of our spin measurements.


\subsection{Comparison with the Reflection Measurements}

Another widely-used technique of measuring the spin is the iron-line profile fitting method (\citealt{Fabian+etal+1989}, \citealt{Reynolds+etal+2003}). As mentioned above, some portion of the power-law component is reflected by the disk, generating the reflection spectrum whose main prominent features  include the fluorescent emission lines (in particularly, the iron K$\alpha$ line between 6.4 - 6.97 keV, depending on ionization state) and the Compton hump at 20 - 30 keV (\citealt{Ross+etal+1999}). The intrinsically narrow iron K$\alpha$ line is then broadened and smeared by the combined effects of Doppler shift, special relativity, and general relativity (\citealt{Fabian+etal+2000}), and by fitting the iron-line profile, one can deduce the spin parameter.

The spin results from the continuum-fitting method both in the previous work and in this work are consistent with the spins derived from the reflection method (\citealt{Duro+etal+2011}, \citealt{Fabian+etal+2012}, \citealt{Tomsick+etal+2014}, \citealt{Parker+etal+2015}, \citealt{Duro+etal+2016}, \citealt{Walton+etal+2016}, \citealt{Tomsick+etal+2018}), although \cite{Miller+etal+2002} found a near-zero spin of Cygnus X-1, which is mainly because the lack of adequate energy coverage ($\lesssim$ 10 keV) made it difficult to subtract the underlying continuum properly.

It should be noted that the highly-sensitive {\it NuSTAR}, launched in 2012, provides unprecedentedly precise measurements of the iron line region and reflection component. In the soft state, \cite{Tomsick+etal+2014} reported a rapid spin of $a_*$ $>$ 0.83 and \cite{Walton+etal+2016} estimated a higher spin of 0.93 $\lesssim a_* \lesssim$ 0.96. \cite{Parker+etal+2015} analyzed observations taken in the hard state and found a high spin of $a_*$ $>$ 0.97. In the intermediate state, \cite{Tomsick+etal+2018} restricted the spin to be $a_* > 0.93$ for the high-density reflection model. 

In addition, \cite{Duro+etal+2016} analyzed a spectra of four simultaneous hard intermediate state observations, and found a spin of $\simeq$ 0.9. They also suggested that the different inclination angles inferred in the different states (i.e., the reported inclination angle of $\simeq$ 30$^\circ$ in the hard intermediate state, as compared to the higher inclinations found from the reflection measurements in the soft state) could imply a warp in the inner region of the disk.  However, as discussed by \cite{Tomsick+etal+2018}, these results can be somewhat degenerate with the iron abundance and with the disk electron density.

\subsection{Spin Comparison with Other Systems}

We found that the spin of Cygnus X-1 is constrained to be $a_{*} > 0.9985$, which is the highest among all the stellar-mass black hole systems with a measured spin parameter. Although it may be less extreme due to some systematics, this value would be still very high. If we look through the spin parameters measured for the stellar-mass black hole systems using the continuum-fitting method, it shows that almost all the black holes in high-mass X-ray binaries have a relatively high spin (Cygnus X-1: $a_{*} > 0.9985$, this work; LMC X-1: $a_{*}=0.92^{+0.05} _{-0.07}$,~\citealt{Gou+etal+2009}; M33 X-7: $a_{*} = 0.84 \pm 0.05$,~\citealt{Liu+etal+2010}; IC10 X-1: $a_{*} \gtrsim 0.8$,~\citealt{Steiner+etal+2016}). However, the low mass X-ray binaries have a wide range of spins between zero and unity, which may imply different evolutionary paths for these systems (\citealt{Bambi+etal+2019}, \citealt{Reynolds+etal+2019}). 

It is noted that, since the first detection of GW150914 in 2015, LIGO and Virgo have made 3 runs, and also made 50 detections of gravitational wave events with majority of stellar mass binary black hole systems. The results from the first 2 runs have been catalogued in \cite{Abbott+etal+2019}, which shows the effective spin $\chi_{\rm eff}$ \footnote{${\chi_{\rm eff} = \frac {(m_{1}{\vec a_1}+m_{2}{\vec a_2}) \cdot \hat{L_{\rm N}}}{(m_{1}+m_{2})},}$ where $m_{\rm i}$ and $\vec a_{\rm i}$ indicate masses and dimensionless spins of each component, $\hat{L_{\rm N}}$ is the unit vector along the direction of the orbital angular momentum. }of the binary system before merger is quite small with the distribution’s mean $\mu\approx0$, and with a narrow distribution, which is confirmed by \cite{Miller+etal+2020}. \cite{Miller+etal+2020} further explored what these ensemble properties imply about the spins of individual binary black hole mergers, re-analyzing existing gravitational-wave events with a population-informed prior on their effective spin, and found, under this analysis, the binary black hole GW170729, which previously excluded $\chi_{\rm eff}$ = 0, is now consistent with zero effective spin at $\approx10\%$ credibility.  

As we know, the companion star of Cygnus X-1 is around 40 $M_{\odot}$, and based on our understanding to the stellar evolution, it may eventually produce a second black hole of the system, forming a binary black hole system in the future. In addition, the current analysis to the system shows that the black hole spin-orbit misalignment is small, within 10 degree (\citealt{Miller-Jones+etal+2020}). If the black hole of Cygnus X-1 and the companion star were formed from the same cloud initially, then the companion star is usually aligned with the orbital plane. In this case, we should expect the effective spin of the system is quite large, unless there is a large asymmetric kick during the black hole formation of the companion star, and it will be quite different from the spin distribution obtained from the GW results. This may imply that Cygnus X-1 system is likely to follow a different formation channel (primordial one) compared to the GW black hole system, although it is not clear whether the GW results preferentially support a dynamical or field binary origin (or even other possibilities) for stellar mass binary black holes detected by GWs. Furthermore, \cite{Miller-Jones+etal+2020} shows that, given the current orbital separation, we do not expect Cygnus X-1 or systems with similar properties to yield binary black hole mergers in the age of the Universe. So we may haven't detect any GW signal from primordial binary black hole systems yet. 

As to the origin of the high spin in these high-mass systems, it is usually concluded that the spin is natal (\citealt{Podsiadlowski+etal+2003}, \citealt{Axelsson+etal+2011}, \citealt{Qin+etal+2019}), at least for those high-mass X-ray binaries, mainly because in the whole lifetime of a stellar-mass black hole, it is centainly impossible to spin up from zero to a highly-spinning black hole via standard (as opposed to hypercritical) accretion (\citealt{Wong+etal+2012}). One possible evolutionary channel to explain the high spin is that the BH progenitor undergoes mass transfers while still in the main sequence (see Case-A MT in \citealt{Qin+etal+2019}). Hence the core of the BH progenitor contains enough angular momentum to form a highly rotating black hole (many observational features predicted by this path, such as unusual surface abundances of the secondary, have been found in Cygnus X-1, as mentioned in \citealt{Miller-Jones+etal+2020}).


\section{conclusion} \label{conc}

We re-analyze six archival spectra of the black hole X-ray binary Cygnus X-1, which were originally presented in GOU11 and GOU14, to constrain the black hole spin. We adopt up-to-date values of three key dynamical parameters $M$, $i$, and $D$. These data rigorously satisfy the selection criterion f$_{\rm sc} \lesssim 25\%$. The key model we utilized is a fully relativistic thin disk model {\sc kerrbb{\small 2}}. Monte Carlo simulations are performed in order to estimate the error in $a_{*}$ due to the combined uncertainties of $M$, $i$, and $D$. Ultimately, we arrive at our final result, $a_{*} > 0.9985$ at the 3$\sigma$ level of confidence, which confirms the extreme spin of the black hole in Cygnus X-1. 

It should be noted that the spin value we presented above is the theoretical one purely based on the model itself, and the hard limit is set at 0.9999. In practice, the maximum spin value could be slightly lower than 0.9999. \cite{Thorne+1974} estimated the evolution of a black hole, and pointed out that the maximum spin of a black hole can only reach 0.998 because the radiation from the accretion disk will produce a counteracting torque on the black hole once the black hole reaches the maximum spin. Given the Thorne limit, we conclude that the black hole in Cygnus X-1 has an extreme spin in any event. 

\acknowledgments

We thank the referee for his/her constructive suggestions. LJG acknowledges the support by the National Program on Key Research and Development Project through grant No. 2016YFA0400804, and by the NSFC with grant No. U1838114, and by the Strategic Priority Research Program of the Chinese Academy of Sciences through grant No. XDB23040100.  JCAM-J is the recipient of an Australian Research Council Future Fellowship (FT140101082), funded by the Australian government.


\begin{table}
\label{tab1}
\caption{Log of the Observations$^a$}
\scalebox{0.8}{
\begin{threeparttable}
\footnotesize
\begin{tabular}{ccccccccc}
\toprule
\toprule
\specialrule{0em}{1pt}{1.3pt}
Number&ObsID&Mission&Detector&Energy Band (keV)&UT&T$_{\rm exp}$ (s)&I (Crab)&$\phi$\\
\specialrule{0em}{1pt}{1.3pt}
\midrule
\specialrule{0em}{1pt}{1.3pt}  
SP1&10408000 $\&$ P10412&{\it ASCA} $\&$ {\it RXTE}&GIS $\&$ PCA&0.7–8.0 $\&$ 2.5–45.0&1996-05-30 06:43:16 $\&$ 07:51:29&2547 $\&$ 2240&0.80&0.74\\
\specialrule{0em}{1pt}{1.3pt} 
SP2&12472 $\&$ P96378&{\it Chandra} $\&$ {\it RXTE}&HETG(CC) $\&$ PCA&0.8–8.0 $\&$ 2.9–50&2011-01-06 14:06:40–14:35:44&455 $\&$ 1744&0.52&0.32\\
\specialrule{0em}{1pt}{1.3pt} 
SP3&12472 $\&$ P96378&{\it Chandra} $\&$ {\it RXTE}&HETG(CC) $\&$ PCA&0.8–8.0 $\&$ 2.9–50&2011-01-06 15:44:16–16:09:52&398 $\&$ 1536&0.61&0.33\\
\specialrule{0em}{1pt}{1.3pt} 
SP4&12472 $\&$ P96378&{\it Chandra} $\&$ {\it RXTE}&HETG(CC) $\&$ PCA&0.8–8.0 $\&$ 2.9–50&2011-01-06 17:15:28–17:43:44&455 $\&$ 1744&0.57&0.35\\
\specialrule{0em}{1pt}{1.3pt} 
SP5&12472 $\&$ P96378&{\it Chandra} $\&$ {\it RXTE}&HETG(CC) $\&$ PCA&0.8–8.0 $\&$ 2.9–50&2011-01-06 18:19:44–19:17:52&455 $\&$ 1744&0.38&0.36\\
\specialrule{0em}{1pt}{1.3pt} 
SP6&12472 $\&$ P96378&{\it Chandra} $\&$ {\it RXTE}&HETG(CC) $\&$ PCA&0.8–8.0 $\&$ 2.9–50&2011-01-06 19:53:36–20:50:08&455 $\&$ 1744&0.38&0.37\\
\specialrule{0em}{1pt}{1.3pt}
\bottomrule
\end{tabular}
\end{threeparttable}}

\tablecomments{$^a$ The log of the observations analyzed in this work (adapted from observing logs in GOU11, GOU14). In columns 2--9, we show the following information, respectively: observation identifier; names of the observational missions; names of the detectors requested for the exposure; energy band used for the spectral fit; the start and end times of the observations; the effective exposure times; the intensity of the source; The orbital phase of the binary system, which is defined (at the midpoint of the observation) relative to the time of superior conjunction of the O-star (black hole beyond star), occurring on heliocentric Julian Day 2441163.529~\citep{Orosz+etal+2011}.
}

\end{table}


\begin{table}
\label{tab2}
\caption{Best-fit Results for Our Adopted Model$^a$}
\scalebox{0.65}{
\begin{threeparttable}
\begin{tabular}{cccccccccc}
\toprule
\toprule
\specialrule{0em}{1pt}{1.3pt}
\multirow{2}{*}{Number}&\multirow{2}{*}{Model}&\multirow{2}{*}{Parameter$^a$}
&\multicolumn{6}{c}{Spectral Number}&\multirow{2}{*}{Case 1}$^b$\\
\cmidrule{4-9}
&&&SP1&SP2&SP3&SP4&SP5&SP6\\
\specialrule{0em}{1pt}{1.3pt}
\midrule
\specialrule{0em}{1pt}{1.3pt}  
\specialrule{0em}{1pt}{1.3pt}  
1 & {\sc tbabs} & $N_{\rm H}$ & 0.74 $\pm$ 0.01 & 0.82 $\pm$ 0.02 & 0.83 $\pm$ 0.03 & 0.80 $\pm$ 0.02 & 0.79 $\pm$ 0.02 & 0.81 $\pm$ 0.01 & 0.72 $\pm$ 0.01\\
\specialrule{0em}{1pt}{1.3pt}
2 & {\sc simplr} & $\Gamma$ & 2.39 $\pm$ 0.01 & 2.38 $\pm$ 0.04 & 2.46 $\pm$ 0.03 & 2.69 $\pm$ 0.02 & 2.62 $\pm$ 0.03 & 2.50 $\pm$ 0.01 & 2.45 $\pm$ 0.01\\
\specialrule{0em}{1pt}{1.3pt}
3 & {\sc simplr} & $f_{\rm sc}$ & 0.247 $\pm$ 0.005 & 0.106 $\pm$ 0.016 & 0.156 $\pm$ 0.012 & 0.211 $\pm$ 0.010 & 0.163 $\pm$ 0.009 & 0.093 $\pm$ 0.002 & 0.092 $\pm$ 0.002\\
\specialrule{0em}{1pt}{1.3pt}
4 & {\sc kerrbb2} & $a_*$ & $0.9999 ^{+0.0000}_{-0.0087}$ & $0.9999 ^{+0.0000}_{-0.0137}$ & $0.9999 ^{+0.0000}_{-0.0155}$ & $0.9999 ^{+0.0000}_{-0.0130}$ & $0.9999 ^{+0.0000}_{-0.0115}$ & $0.9999 ^{+0.0000}_{-0.0015}$ & $0.9696 \pm 0.0034$ \\
\specialrule{0em}{1pt}{1.3pt}
5 & {\sc kerrbb2} & $\dot{M}$ & 0.17 $\pm$ 0.02 & 0.20 $\pm$ 0.03 & 0.19 $\pm$ 0.04 & 0.19 $\pm$ 0.03 & 0.18 $\pm$ 0.03 & 0.18 $\pm$ 0.01 & 0.23 $\pm$ 0.01\\
\specialrule{0em}{1pt}{1.3pt}
6 & {\sc kerrdisk} & $E_{\rm L}$ & 6.57 $\pm$ 0.06 & 6.74 $\pm$ 0.08 & 6.55 $\pm$ 0.07 & 6.40 $\pm$ 0.13 & 6.55 $\pm$ 0.11 & 6.70 $\pm$ 0.08 & 6.46 $\pm$ 0.06\\
\specialrule{0em}{1pt}{1.3pt}
7 & {\sc kerrdisk} & $q$ & 2.89 $\pm$ 0.08 & 3.00 $\pm$ 0.09 & 2.73 $\pm$ 0.11 & 1.64 $\pm$ 0.40 & 2.66 $\pm$ 0.17 & 2.95 $\pm$ 0.06 & 3.45 $\pm$ 0.10\\
\specialrule{0em}{1pt}{1.3pt}
8 & {\sc kerrdisk} & $N_{\rm L}$ & 0.009 $\pm$ 0.001 & 0.024 $\pm$ 0.003 & 0.029 $\pm$ 0.003 & 0.012 $\pm$ 0.002 & 0.011 $\pm$ 0.002 & 0.014 $\pm$ 0.001 & 0.026 $\pm$ 0.001\\
\specialrule{0em}{1pt}{1.3pt}
9 & {\sc kerrdisk} & EW & 0.11 & 0.36 & 0.29 & 0.14 & 0.21 & 0.29 & 0.61\\
\specialrule{0em}{1pt}{1.3pt}
10 & {\sc ireflect} & $X_{\rm Fe}$ & 6.3 $\pm$ 0.3 & 6.0 $\pm$ 1.1 & 3.9 $\pm$ 0.5 & 3.5 $\pm$ 0.3 & 5.9 $\pm$ 0.6 & 4.1 $\pm$ 0.6 & 3.5 $\pm$ 0.6\\
\specialrule{0em}{1pt}{1.3pt}
11 & {\sc ireflect} & $\xi$ & 62 $\pm$ 10 & 335 $\pm$ 186 & 137 $\pm$ 45 & 21 $\pm$ 7 & 57 $\pm$ 18 & 381 $\pm$ 51 & 412 $\pm$ 64\\
\specialrule{0em}{1pt}{1.3pt}
12 & {\sc constant} & $N_{\rm ASCA}$ & 0.955 $\pm$ 0.003 & - & - & - & - & -&-\\
\specialrule{0em}{1pt}{1.3pt}
13 & {\sc constant} & $N_{\rm MEG}$ & - & 0.688 $\pm$ 0.009 & 0.575 $\pm$ 0.007 & 0.704 $\pm$ 0.007 & 0.713 $\pm$ 0.005 & 0.682 $\pm$ 0.006 & 0.690 $\pm$ 0.011\\
\specialrule{0em}{1pt}{1.3pt}
14&{\sc constant}&$N_{\rm HEG}$ & - & 0.863 $\pm$ 0.011 & 0.889 $\pm$ 0.012 & 0.887 $\pm$ 0.009 & 0.923 $\pm$ 0.007 & 0.889 $\pm$ 0.008 & 0.903 $\pm$ 0.014\\
\specialrule{0em}{1pt}{1.3pt}
\midrule
\specialrule{0em}{1pt}{1.3pt}
15 & & Reduced $\chi^2$ ($\chi^2$/degree)& 1.49 (678.09/454)& 1.12 (705.15/627)& 1.15 (656.82/572)& 1.32 (823.61/624)& 1.25 (929.81/744)& 1.24 (1241.40/998)& 1.09 (1087.79/998)\\
\specialrule{0em}{1pt}{1.3pt}
16 & & $f$ & 1.6 & 1.6 & 1.6 & 1.6 & 1.6 & 1.6 & 1.5\\
\specialrule{0em}{1pt}{1.3pt}
17 & & $l$ & 0.02 & 0.03 & 0.03 & 0.02 & 0.02 & 0.02 & 0.02\\
\bottomrule
\end{tabular}
\end{threeparttable}}
\tablecomments{$^a$ Names of the different model components are listed in column 2. In column 3, the parameters from top to bottom are: (1) equivalent hydrogen column density $N_{\rm H}$ in units of 10$^{22}$ atoms cm$^{-2}$; (2) the photon power-law index $\Gamma$; (3) the scattered fraction $f_{\rm sc}$; (4) the dimensionless spin parameter $a_{*}$ (the upper limit is set to 0.9999); (5) the mass accretion rate $\dot{M}$ in units of 10$^{18}$ g s$^{-1}$; (6) the rest frame central line energy $E_{\rm L}$, in units of keV; (7) the emissivity index $q$; (8) the line flux $N_{\rm L}$ in units of photons\,cm$^{-2}$\,s$^{-1}$; (9) the equivalent width EW of the line in units of keV; (10) the iron abundance $X_{\rm Fe}$, relative to the solar abundance; (11) the disk ionization parameter $\xi$ in units of erg\,cm\,s$^{-1}$; (12) the detector normalization constant $N_{\rm ASCA}$ of {\it ASCA} relative to {\it RXTE} PCU2; (13) the detector normalization constant $N_{\rm MEG}$ of {\it Chandra} MEG relative to {\it RXTE} PCU2; (14) the detector normalization constant $N_{\rm HEG}$ of {\it Chandra} HEG relative to {\it RXTE} PCU2; (15) the reduced chi-squared value, total chi-squared value and number degrees of freedom; (16) the spectral hardening factor $f$; (17) the bolometric, Eddington-scaled X-ray luminosity of the thermal disk component, $l$ = $L$/$L_{\rm Edd}$, where $L$ = $L(a_*,\dot M)$ and $L_{\rm Edd}$ = 2.8 $\times 10^{39}$ erg $s^{-1}$ for $M$ = 21.2 $M\odot$. The uncertainties are quoted at the 90\% confidence level.\\
$^b$Case 1: fit to SP6 with the inclination angle $i$ fixed at $42.47^\circ$. The parameter configuration is exactly the same as that in column 5-9.
}

\end{table}


\begin{table}
\label{tab3}
\caption{Best-fit Results for a Hybrid Thermal/Non-thermal Comptonization Model}
\scalebox{0.65}{
\begin{threeparttable}
\begin{tabular}{cccccccccc}
\toprule
\toprule
\specialrule{0em}{1pt}{1.3pt}
\multirow{2}{*}{Number}&\multirow{2}{*}{Model}&\multirow{2}{*}{Parameter}
&\multicolumn{6}{c}{Spectral Number}&\multirow{2}{*}{Case 1}$^b$\\
\cmidrule{4-9}
&&&SP1&SP2&SP3&SP4&SP5&SP6\\
\specialrule{0em}{1pt}{1.3pt}
\midrule
\specialrule{0em}{1pt}{1.3pt}  
\specialrule{0em}{1pt}{1.3pt}  
1 & {\sc tbabs} & $N_{\rm H}$ & 0.73$\pm$0.01 & 0.73 $\pm$ 0.05 & 0.71 $\pm$ 0.06 & 0.65 $\pm$ 0.06 & 0.71 $\pm$ 0.03 & 0.75 $\pm$ 0.04& 0.76 $\pm$ 0.02\\
\specialrule{0em}{1pt}{1.3pt}
2 & {\sc kerrbb2} & $a_*$ & 0.9947 $\pm$ 0.0014 & $0.9999 ^{+0.0000}_{-0.1587}$ &  $0.9999 ^{+0.0000}_{-0.2105}$ & $0.9999 ^{+0.0000}_{-0.1798}$ & $0.9999 ^{+0.0000}_{-0.0903}$ & $0.9999 ^{+0.0000}_{-0.1449}$ &$0.9999 ^{+0.0000}_{-0.0169}$\\
\specialrule{0em}{1pt}{1.3pt}
3 & {\sc kerrbb2} & $\dot{M}$ & 0.18 $\pm$ 0.01 & 0.14 $\pm$ 0.06 & 0.22 $\pm$ 0.02 & 0.25 $\pm$ 0.03 & 0.13 $\pm$ 0.10 & 0.29 $\pm$ 0.05 & 0.16 $\pm$ 0.03\\
\specialrule{0em}{1pt}{1.3pt}
4 & {\sc simplr} & $\Gamma$ & 1.99 $\pm$ 0.01 & 2.40 $\pm$ 0.04 & 2.46 $\pm$ 0.02 & 2.53 $\pm$ 0.02 & 2.43 $\pm$ 0.05 & 2.52 $\pm$ 0.02 & 2.39 $\pm$ 0.02\\
\specialrule{0em}{1pt}{1.3pt}
5 & {\sc simplr} & $f_{\rm sc}$ & 0.399 $\pm$ 0.007 & 0.447 $\pm$ 0.440 & 0.431 $\pm$ 0.359 & 0.393 $\pm$ 0.198 & 0.361 $\pm$ 0.234 & 0.361 $\pm$ 0.311 & 0.492 $\pm$ 0.176\\
\specialrule{0em}{1pt}{1.3pt}
6 & {\sc comptt} & $T_{\rm 0}$ $^a$ & 0.45$\pm$0.06  & 0.45 $\pm$ 0.44 & 0.40 $\pm$ 0.23 & 0.41 $\pm$ 0.36 & 0.39 $\pm$ 0.16 & 0.42 $\pm$ 0.26 &0.44(fixed)\\
\specialrule{0em}{1pt}{1.3pt}
7 & {\sc comptt} & $kT$ $^a$ & 5.77$\pm$2.51  & $3.87 ^{+125.44}_{-3.77}$ & $4.00 ^{+113.39}_{-3.90}$ & $4.11 ^{+164.92}_{-4.01}$ & $4.06 ^{+96.8}_{-3.96}$ & $3.95 ^{+109.05}_{-3.85}$&3.4(fixed)\\
\specialrule{0em}{1pt}{1.3pt}
8 & {\sc comptt} & $\tau$ $^a$  & 2.08$\pm$0.74  &$0.81 ^{+37.60}_{-0.80}$ & $0.63 ^{+26.64}_{-0.62}$ &$0.33 ^{+23.50}_{-0.32}$ & $0.80 ^{+26.65}_{-0.79}$ & $0.43 ^{+19.5}_{-0.42}$ &1.66(fixed)\\
\specialrule{0em}{1pt}{1.3pt}
9 & {\sc comptt} & $\rm Norm$ $^a$ & 1.08$\pm$0.48  & $2.50 ^{+88.45}_{-2.50}$  & $3.64 ^{+110.73}_{-3.64}$ & $3.86 ^{+165.12}_{-3.86}$ & $2.15 ^{+55.37}_{-2.15}$ & $2.41 ^{+70.82}_{-2.41}$ &2.12$\pm$ 0.07\\
\specialrule{0em}{1pt}{1.3pt}
10 & {\sc kerrdisk} & $E_{\rm L}$ &  6.59$\pm$0.04 & 6.56 $\pm$ 0.07 & 6.51 $\pm$ 0.08 & 6.49 $\pm$ 0.08 & 6.57 $\pm$ 0.07 & 6.55 $\pm$ 0.08& 6.54 $\pm$ 0.07\\
\specialrule{0em}{1pt}{1.3pt}
11 & {\sc kerrdisk} & $q$ & 2.74$\pm$0.03  & 2.51 $\pm$ 0.19 & 2.50 $\pm$ 0.21 & 2.52 $\pm$ 0.22 & 2.53 $\pm$ 0.19 & 2.57 $\pm$ 0.19& 2.38 $\pm$ 0.16\\
\specialrule{0em}{1pt}{1.3pt}
12 & {\sc kerrdisk} & $N_{\rm L}$ & 0.019$\pm$0.001  & 0.020 $\pm$ 0.004 & 0.023 $\pm$ 0.004 & 0.020 $\pm$ 0.004 & 0.014 $\pm$ 0.003 & 0.013 $\pm$ 0.002& 0.019 $\pm$ 0.002 \\
\specialrule{0em}{1pt}{1.3pt}
13 & {\sc kerrdisk} & EW & 0.21  & 0.29 & 0.25 & 0.23 & 0.29 & 0.26& 0.26\\
\specialrule{0em}{1pt}{1.3pt}
14 & {\sc ireflect} & $X_{\rm Fe}$ & 2.2$\pm$0.1 & 5.3 $\pm$ 1.2 & 3.8 $\pm$ 0.6 & 4.5 $\pm$ 0.8 & 4.1 $\pm$ 1.0 & 3.9 $\pm$ 0.8 & 4.4 $\pm$ 0.5\\
\specialrule{0em}{1pt}{1.3pt}
15 & {\sc ireflect} & $\xi$ & 198$\pm$11  & 107 $\pm$ 48 & 80 $\pm$ 31 & 92 $\pm$ 37 & 111 $\pm$ 48 & 128 $\pm$ 53 &89 $\pm$ 19 \\
\specialrule{0em}{1pt}{1.3pt}
16 & {\sc constant} & $N_{\rm ASCA}$ & 1.012$\pm$0.003  & - & - & - & - & -& -\\
\specialrule{0em}{1pt}{1.3pt}
17 & {\sc constant} & $N_{\rm MEG}$ & - & 0.824 $\pm$ 0.010 & 0.664 $\pm$ 0.008 & 0.813 $\pm$ 0.009 & 0.803 $\pm$ 0.007 & 0.800 $\pm$ 0.008& 0.820 $\pm$ 0.009\\
\specialrule{0em}{1pt}{1.3pt}
18 &{\sc constant}&$N_{\rm HEG}$ & - & 1.037 $\pm$ 0.012 & 1.029 $\pm$ 0.012 & 1.026 $\pm$ 0.012 & 1.039 $\pm$ 0.009 & 1.045 $\pm$ 0.010 & 1.031 $\pm$ 0.011\\
\specialrule{0em}{1pt}{1.3pt}
\midrule
\specialrule{0em}{1pt}{1.3pt}
19 & & Reduced $\chi^2$ ($\chi^2$/degree)& 0.94 (430.62/451) & 0.93 (581.00/624)& 1.02 (580.58/572)& 0.94 (582.79/621)& 1.27 (941.07/741)& 1.12 (1114.26/994)& 0.93 (583.43/627)\\
\specialrule{0em}{1pt}{1.3pt}
20 & & $f$ & 1.6  & 1.6 & 1.5 & 1.6 & 1.6 & 1.6& 1.6\\
\specialrule{0em}{1pt}{1.3pt}
21 & & $l$ & 0.02  & 0.02 & 0.01 & 0.02 & 0.02 & 0.02& 0.02\\
\bottomrule
\end{tabular}
\end{threeparttable}}
\tablecomments{$^a$ Names of different parameters in model {\sc comptt} are (1) the input soft photon Wien temperature $T_{\rm 0}$ in units of keV; (2) the plasma temperature $kT$ in units of keV (the lower limit is set to 0.1 keV); (3) the plasma optical depth $\tau$ (the lower limit is set to 0.01); (4) the normalization $norm$ (the lower limit is set to 0).\\
$^b$Case 1: fit to SP2 with $T_{\rm 0}$ fixed at 0.44 keV, $kT$ fixed at 3.4 keV and $\tau$ fixed at 1.66. The parameter configuration is exactly the same as that in column 5-9.
}

\end{table}

\begin{figure}
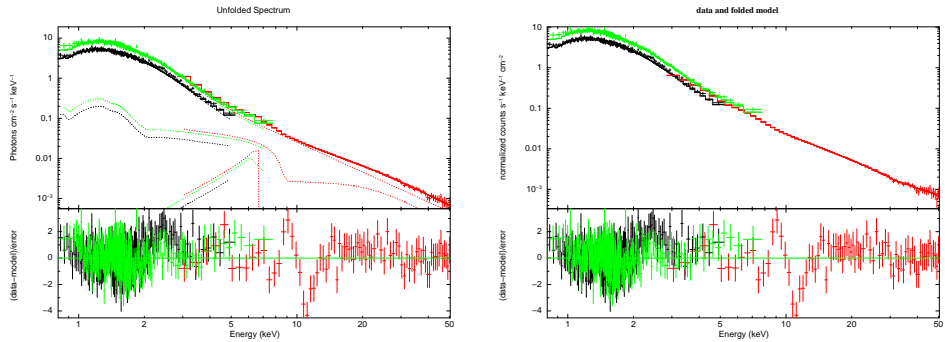

    \epsscale{1.0}
    \centering
    \subfigure{
    \includegraphics[width=0.25\textwidth,angle=270]{2_uf_del.eps}}
    \subfigure{
    \includegraphics[width=0.25\textwidth,angle=270]{lda_del_2.eps}}
    \caption{A representative best-fit X-ray spectrum (SP3). Left: A plot of the unfolded spectrum and the model. Right: A plot of the spectrum with the folded model. Detailed fit parameters are listed in Table~2. The data obtained by {\it RXTE} is plotted in red. For the {\it Chandra} observation we used, the data collected by the High Energy Grating (HEG) and Medium Energy Grating (MEG) components of the High-Energy Transmission Grating (HETG) are plotted as green and black, respectively. The dotted lines represent the models we adopted.}
\label{fig1}

\end{figure}


\begin{figure} 
    \epsscale{1.0}
    \plotone{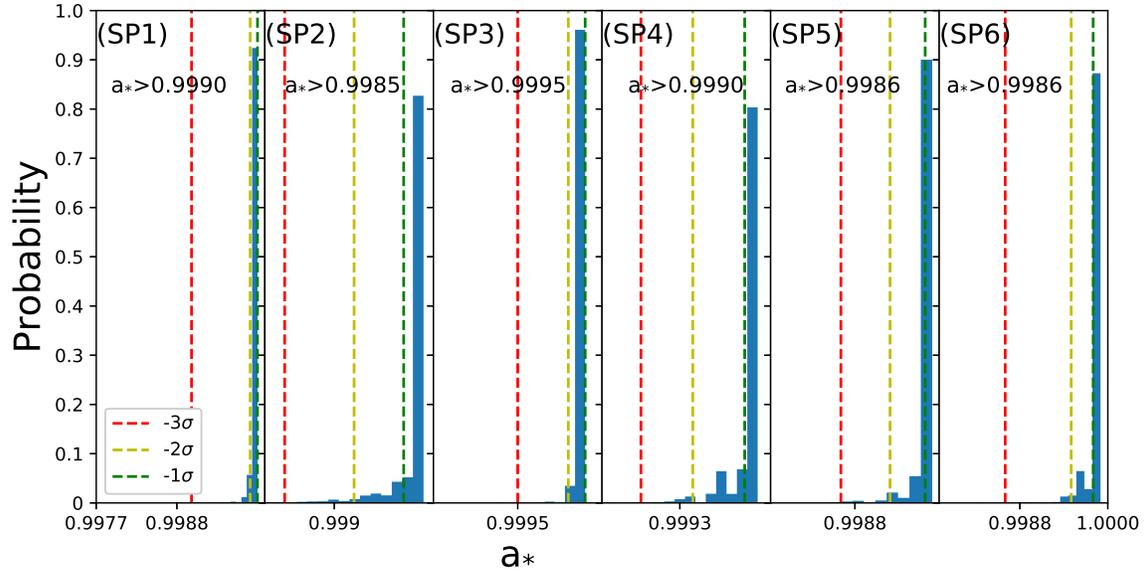}
    \caption{Histograms of $a_{*}$ calculated via the Monte Carlo analysis for 3000 sets of parameters per spectrum. The three dashed lines from left to right imply the 99.7\% (3$\sigma$, red), 95.4\% (2$\sigma$, yellow), and 68.3\% (1$\sigma$, green) error, respectively. The respective 99.7\% confidence lower limit on $a_*$ is indicated in each panel.}
    \label{fig2}
\end{figure}


\begin{figure} 
    \epsscale{1.0}
    \plotone{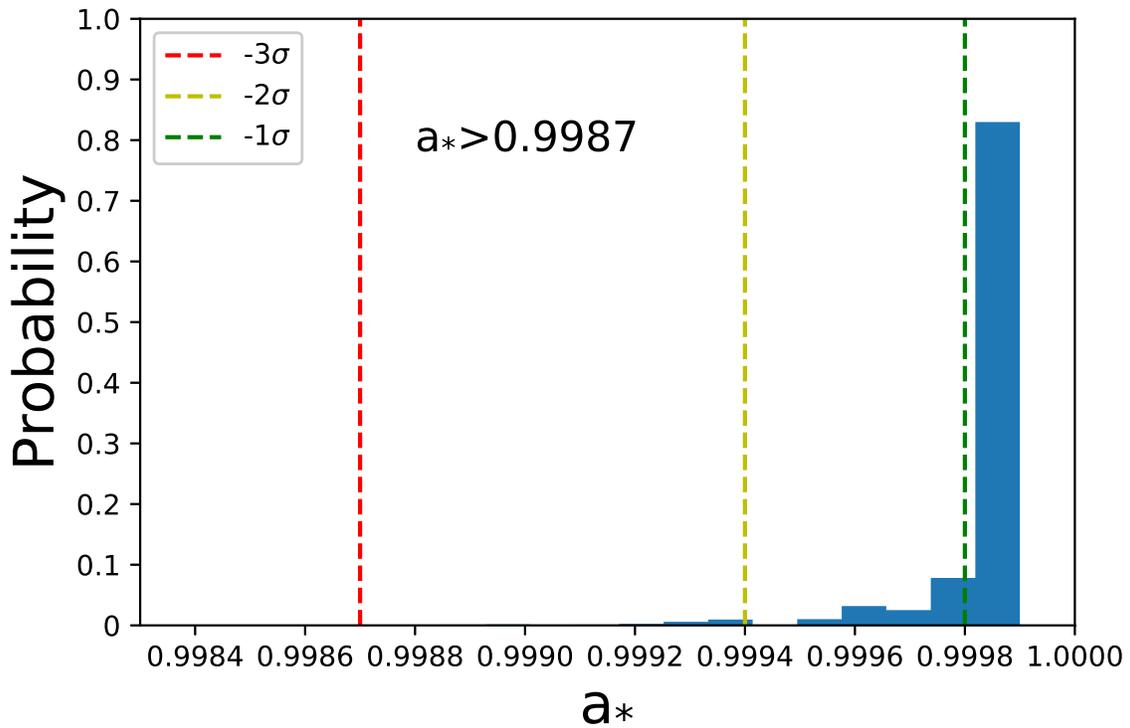}
    \caption{Summed histogram of $a_{*}$ for our six spectra, including 18000 data points. The dashed lines and listed 99.7\% confidence lower limit in this figure are as those shown in Figure~\ref{fig2}.
     }
\label{fig3}
\end{figure}

\begin{figure} 
    \epsscale{1.0}
    \plotone{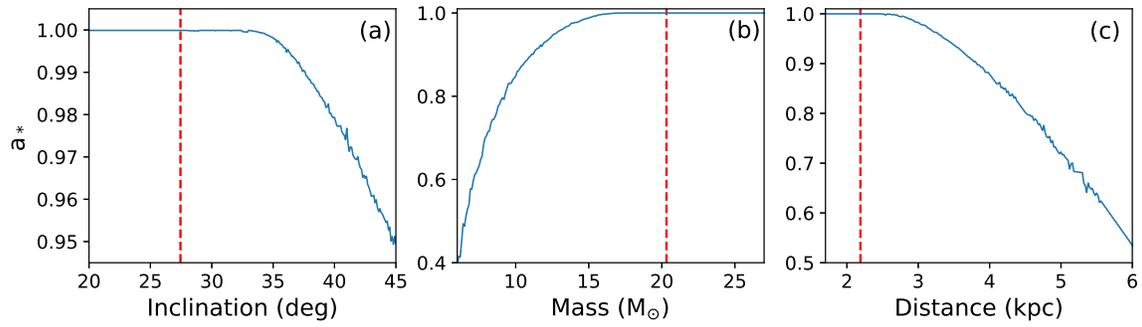}
    \caption{Effects on $a_{*}$ of varying one of the three input parameters while fixing the other two at their updated  values. (a) Spin vs. inclination angle $i$; (b) spin vs. black hole mass $M$; (c) spin vs. source distance $D$. The red dashed lines represent the baseline values $i$ = 27.51 $^{\circ}$, $M$ = 21.2 and $D$ = 2.22 kpc. These plots were made using the fit results of SP6. }
\label{fig4}
\end{figure}

\end{document}